\documentclass[11pt]{article}
\usepackage{epsfig}
\usepackage{latexsym}
\usepackage{graphicx}
\usepackage{lscape}
\begin{document}

\title{Effect of Field Direction and Field Intensity on Directed
Spiral Percolation}
 
\author {Santanu Sinha and S. B. Santra$^*$\\ Department of Physics,
  Indian Institute of Technology Guwahati,\\ Guwahati-781039, Assam,
  India.\\ $^*$santra@iitg.ernet.in}

\date \today

\maketitle

\centerline{\bf Abstract}

Directed spiral percolation (DSP) is a new percolation model with
crossed external bias fields. Since percolation is a model of
disorder, the effect of external bias fields on the properties of
disordered systems can be studied numerically using DSP. In DSP, the
bias fields are an in-plane directional field ($E$) and a field of
rotational nature ($B$) applied perpendicular to the plane of the
lattice. The critical properties of DSP clusters are studied here
varying the direction of $E$ field and intensities of both $E$ and $B$
fields in $2$ dimensions. The system shows interesting and unusual
critical behaviour at the percolation threshold. Not only the
universality class of DSP model is found to belong in a new
universality class than that of other percolation models but also the
universality class remains invariant under the variation of $E$ field
direction. Varying the intensities of the $E$ and $B$ fields, a
crossover from DSP to other percolation models has been studied. A
phase diagram of the percolation models is obtained as a function of
intensities of the bias fields $E$ and $B$.


\section{INTRODUCTION}
Recently, a new site percolation model, directed spiral percolation
(DSP)\cite{dsp,sinha}, was constructed imposing both the directional
and rotational constraints simultaneously on the ordinary percolation
(OP)\cite{prc} model. A new type of anisotropic percolation cluster,
different from directed percolation (DP)\cite{zallen} and spiral
percolation (SP)\cite{rsb}, was found in the DSP model. The model
showed unusual and interesting critical behaviour at the percolation
threshold for the clusters as well as their hulls. It was found that
the DSP model belongs to a new universality class than that of other
percolation models.

There are certain physical properties of a system needed to be studied
in the presence of external biases like electric or magnetic
fields. The electric field gives rise to a directional constraint and
the magnetic field gives rise to a rotational constraint on the motion
of a classical charged particle in a plane when the magnetic field is
applied perpendicular to the plane of motion. Since percolation is a
model of disorder, Hall effect in disordered systems can be studied
using DSP model. Motion of charged particles under crossed electric
and magnetic fields in disordered systems is an important field of
study in recent time\cite{santradf,hlcomb}.

In this paper, the DSP model is studied on the square and triangular
lattices in $2$-dimensions ($2D$) varying the field intensities and
the direction of the fields. As the direction of a field changes, the
number of components of the field on a lattice also changes. The
universality class of DSP model has been verified for different number
of field components. It is found that the universality class of DSP is
independent of the field directions or the number of components of the
directed field. As the intensities of the fields change, the DSP model
is expected to change to OP, DP or SP models at the appropriate field
intensities. Varying the external field intensities, crossover from
DSP to other percolation models has been studied. A complete phase
diagram is obtained for the percolation models studying the DSP model
under variable field intensities.

Below, the DSP model will be developed under variable field directions
and variable field intensities. The effect of change of field
direction will be discussed first. Intensity effect will be considered
later.

\section{DSP Model under Variable Field Direction}
\begin{figure}
\centerline{\hfill \psfig{file=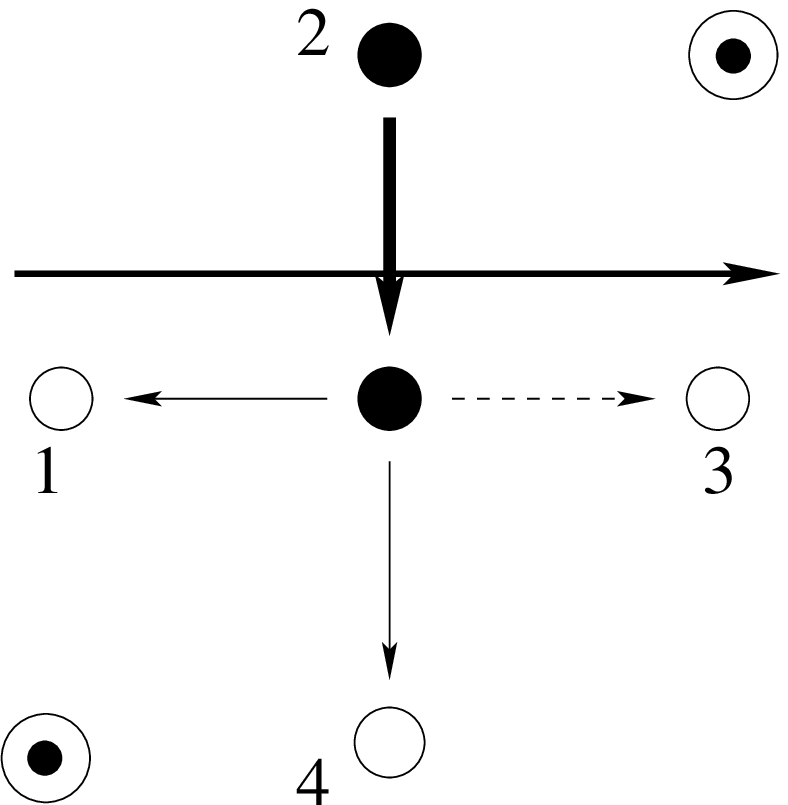,width=0.3\textwidth}
  \hfill\hfill \psfig{file=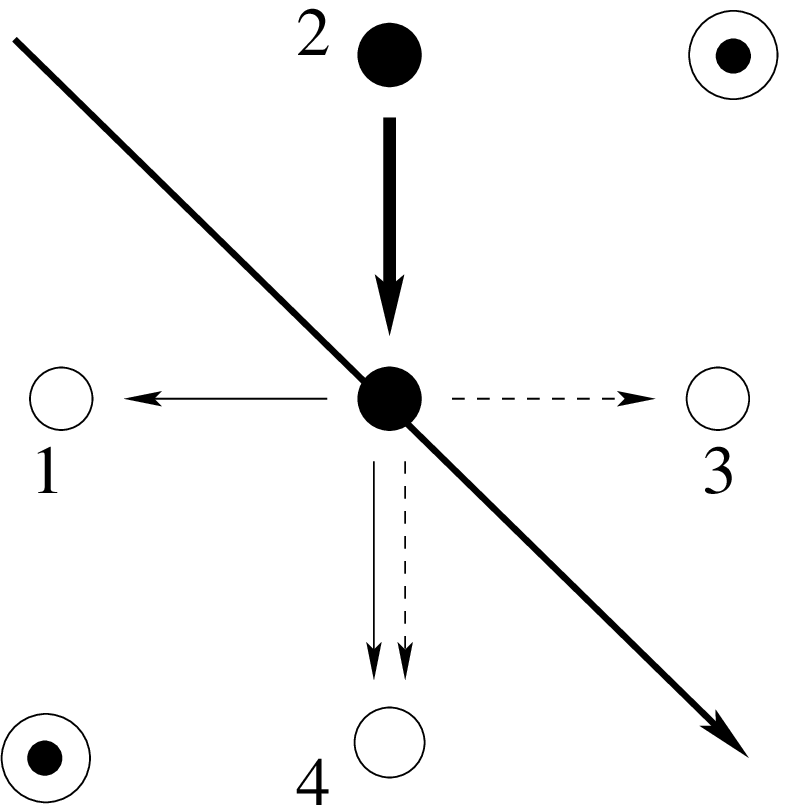,width=0.3\textwidth} \hfill}

\medskip
\centerline{\hfill ($a$) \hfill\hfill ($b$) \hfill}

\bigskip
\bigskip

\centerline{\hfill \psfig{file=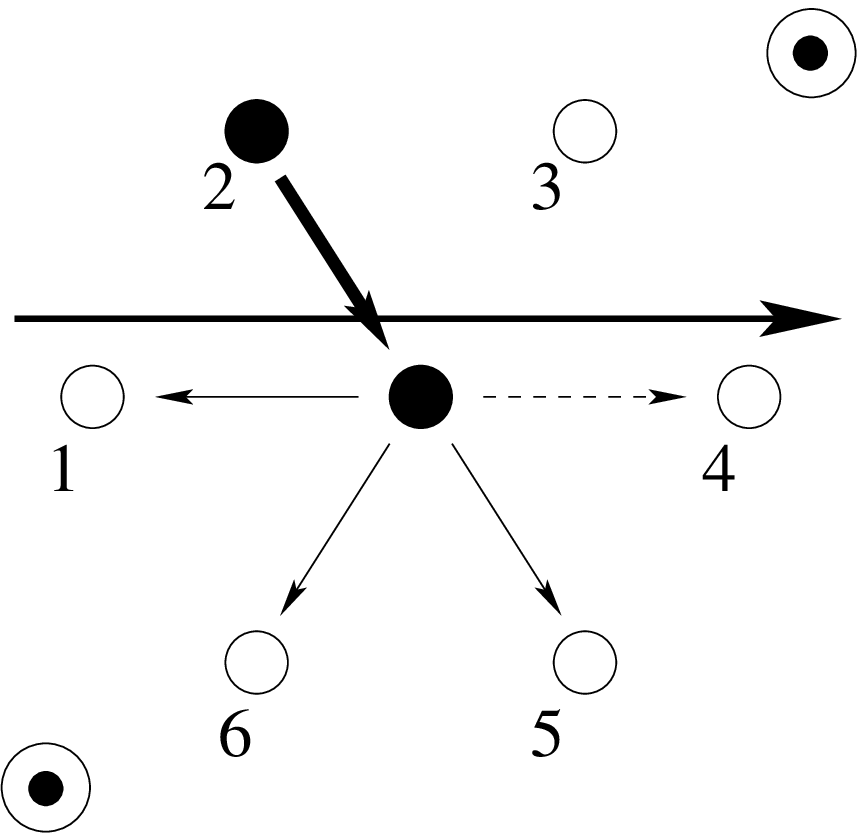,width=0.3\textwidth}
  \hfill\hfill \psfig{file=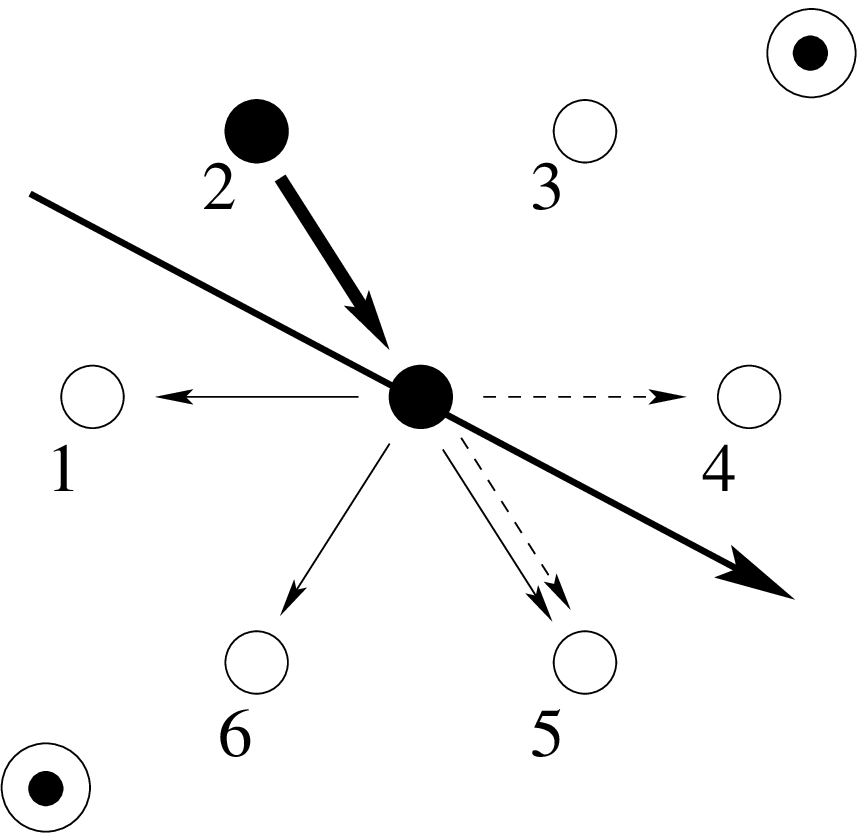,width=0.3\textwidth}\hfill}
 
\medskip
\centerline{\hfill ($c$) \hfill\hfill  ($d$) \hfill}
  
\smallskip
\caption{\label{demo} Selection of empty nearest neighbors for
  occupation in a MC time step on the square ($a$, $b$) and
  triangular ($c$, $d$) lattices. Black circles are the occupied sites
  and the open circles are the empty sites. Thick long arrows
  represent directional constraint ($E$). The clockwise rotational
  constraint ($B$) is shown by encircled dots. The central site is
  occupied from site $2$ and shown by short thick arrows. The eligible
  sites for occupation due to $E$ field are shown by dotted arrows and
  thin solid arrows indicate the same due to $B$ field on both the
  lattices.}
\end{figure}

The DSP model for different field orientations is constructed on the
square and triangular lattices in $2D$. The particles are considered
to be classical charged particles to form percolation clusters under
external bias fields. Two external bias fields perpendicular to each
other are present in the model. An in-plane electric field $E$ gives
rise to a directional constraint and a magnetic field $B$
perpendicular to the plane of the lattice, directed into the plane,
gives rise to clockwise rotational constraint to the occupation of a
lattice site by a particle. A single cluster growth Monte Carlo (MC)
algorithm to generate DSP clusters under different field orientations
will be described here. In this algorithm, the central site of the
lattice is occupied with unit probability. All the nearest neighbors
of the central site can be occupied with equal probability $p$ in the
first time step. As soon as a site is occupied, the direction from
which it was occupied is assigned to it. Once the directional $E$
field is fixed, the empty nearest neighbours in that fixed direction
in space are accessible to occupation due to the directional
constraint. In contrary, for a given $B$ field direction, the empty
nearest neighbours accessible to occupation are in the forward
direction or in a rotational direction, clockwise here, with respect
to the direction of occupation of the present site. Since the forward
and rotational directions depend on the direction of occupation of the
present site, then the rotational constraint is not fixed in
space. First, a list of empty nearest neighbours eligible for
occupation in the next MC time steps is prepared. Selection of
eligible sites on the square lattice is illustrated in
Figs.\ref{demo}($a$) and \ref{demo}($b$) and on the triangular lattice
it is illustrated in Figs.\ref{demo}($c$) and \ref{demo}($d$). On the
square lattice, the $E$ field is in the horizontal direction in
Fig.\ref{demo}($a$) and it is along the upper left to the lower right
diagonal of the lattice in Fig.\ref{demo}($b$). On the triangular
lattice, the $E$ field is horizontal in the Fig.\ref{demo}($c$) and in
Fig.\ref{demo}($d$), the $E$ field is from upper left to the lower
right coroner making an angle $30^\circ$ with the horizontal, a
semi-diagonal field ($E$ field along $60^\circ$ with the horizontal on
the triangular lattice is not considered because it will lead to
anisotropic current distribution). The presence of $E$ field is shown
by the long arrows. The $B$ field is always perpendicular to the plane
of the lattice and directed into the plane as indicated by the
encircled dots. The descriptions given below are valid for all field
configurations given in Fig.\ref{demo}. The black circles represent
the occupied sites and the open circles represent the empty sites. The
direction from which the central site is occupied is represented by a
short thick arrow. In Fig.\ref{demo}, the central site is always
occupied from site $2$. The eligible empty sites for occupation due to
$E$ field are indicated by dotted arrows and thin solid arrows
indicate the same due to $B$ field. The number of eligible empty sites
for occupation due to $B$ field depends on the lattice
structure. There are two such sites on the square lattice, as shown in
Fig.\ref{demo}($a$) and \ref{demo}($b$), whereas there are three such
sites available on the triangular lattice, as shown in
Fig.\ref{demo}($c$) and \ref{demo}($d$) by thin arrows. If the
direction of $B$ field is changed from `into the plane' to `out of the
plane' only the sense of rotational constraint will be changed but the
number of eligible empty sites for occupation will remain the
same. However, the number of eligible empty sites for occupation due
to $E$ field depends on the direction of the $E$ field. There is one
such site for the horizontal $E$ field on both the lattices whereas
there are two such sites for the diagonal $E$ field on the square
lattice and for the semi-diagonal $E$ field on the triangular
lattice. Two eligible sites for occupation in case of diagonal or
semi-diagonal $E$ field orientation is the consequence of two equal
components of the $E$ field along those two nearest neighbours, as
indicated by two dotted arrows in Fig.\ref{demo}(b) and
Fig.\ref{demo}(d).

After selecting the eligible empty sites for occupation, they are
occupied with probability $p$. The coordinate of an occupied site in a
cluster is denoted by $(x$,$y)$. Periodic boundary conditions are
applied in both directions and the coordinates of the occupied sites
are adjusted accordingly whenever the boundary is crossed. At each
time step the span of the cluster in the $x$ and $y$ directions $L_x =
x_{max} - x_{min}$ and $L_y = y_{max} - y_{min}$ are determined. If
either $L_x$ or $L_y$ of a cluster is found greater than or equal to
$L$, the system size, the cluster is then considered as a spanning
cluster. The critical percolation probability $p_c$ is the maximum
probability below which there is no spanning cluster appears and at
$p=p_c$ a spanning cluster appears for the first time in the system.

\section{Effect of Field Direction}

The critical properties of DSP model is studied numerically under
diagonal $E$ field on the square lattice and semi-diagonal $E$ field
on the triangular lattice. For the diagonal or semi-diagonal $E$ field
orientation, there are two components of the field on both the
lattices along which the lattice sites could be occupied. Results
obtained here are compared with that of already obtained results under
horizontal $E$ field orientations, single component of $E$ field, on
the respective lattices\cite{dsp,sinha}. Lattice size is varied from
$L=2^{7}$ to $2^{11}$ in multiple of $2$. Firstly the percolation
thresholds ($p_c$s) are determined for each field configuration on a
given lattice. The cluster related quantities are then evaluated at
their respective $p_c$s. Each data point is averaged over
$N_{tot}=5\times 10^4$ clusters.

\subsection{Percolation threshold}
\begin{figure}
\centerline{\hfill  \psfig{file=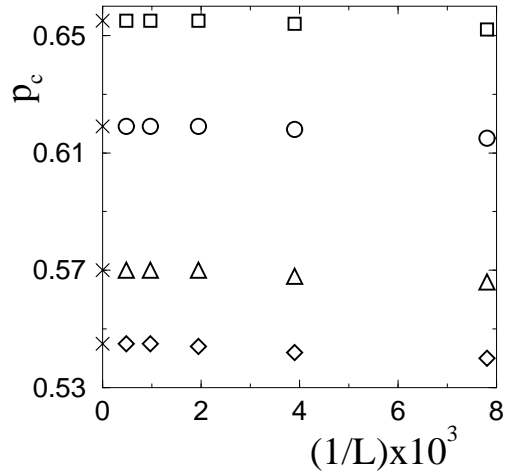,width=0.4\textwidth}\hfill}
\medskip  
\caption{\label{pcl} Plot of critical probability $p_c(L)$ versus
  $1/L$ for different $E$ field orientations. The symbols are: ($\Box$)
  for the horizontal $E$ and ($\bigcirc$) for the diagonal $E$ field on
  the square lattice, ($\triangle$) for the horizontal $E$ and
  ($\Diamond$) for the semi-diagonal $E$ field on the triangular
  lattice. The values of $p_c(L)$ are extrapolated to $L \rightarrow
  \infty$ and $p_c$s for infinite systems are indicated by crosses.  }
\end{figure}

Percolation threshold $p_c$ is the maximum site occupation probability
at which a spanning cluster appears for the first time in the
system. $p_c$s are determined here for the diagonal $E$ field on the
square lattice and semi-diagonal $E$ field on the triangular lattice
in the presence of a crossed $B$ field. The probability to have a
spanning cluster is given by $P_{sp}=n_{sp}/N_{tot}$ where $n_{sp}$ is
the number of spanning clusters out of $N_{tot}$ clusters
generated. The value of $p_c$ for a given system size $L$ and a given
$E$ and $B$ field configuration is determined from the maximum slope
of the curve $P_{sp}$ versus $p$. Generally, $p_c$ depends on $L$ for
finite systems. The value of $p_c$ is then determined as function of
$L$ for a given lattice and field configuration. Data are then
extrapolated to $L\rightarrow \infty$. In Fig.\ref{pcl}, $p_c(L)$ is
plotted against the inverse system size $1/L$ for different field
configurations on both the lattices. Circles represent $p_c$s for the
diagonal $E$ field on the square lattice and diamonds represent the
same for the semi-diagonal $E$ field on the triangular lattice. The
$p_c$s of infinite systems are identified in the $1/L\rightarrow 0$
limit and marked by crosses. It is found that $p_c\approx 0.619$ for
diagonal $E$ field on the square lattice and $\approx 0.545$ for
semi-diagonal $E$ field on the triangular lattice respectively. For
the sake of comparison, $p_c$s for the horizontal field configuration
on both the lattices for different system size are also plotted in the
same figure. The squares represent data for the horizontal $E$ field
on the square lattice and triangles represent the same for the
horizontal $E$ field on the triangular lattice. It was already
obtained that $p_c\approx 0.655$\cite{dsp} and $\approx
0.570$\cite{sinha} for the horizontal $E$ field on the square and
triangular lattices respectively as it is also shown here. As usual,
the value of $p_c$ depends on the type of lattice as well as on the
field situation. If the lattice structure is changed, the number of
components of rotational constraint changes. If the direction of $E$
field changes on a lattice, the number of components of directional
constraint changes. The value of $p_c$ is decreased if the number of
field components is increased either by the rotational constraint or
by the directional constraint. It is expected, because, the clusters
can grow rapidly if the number of field components are higher in a
given lattice.

\subsection{Effective field and Spanning clusters}
\begin{figure}
\centerline{\hfill \psfig{file=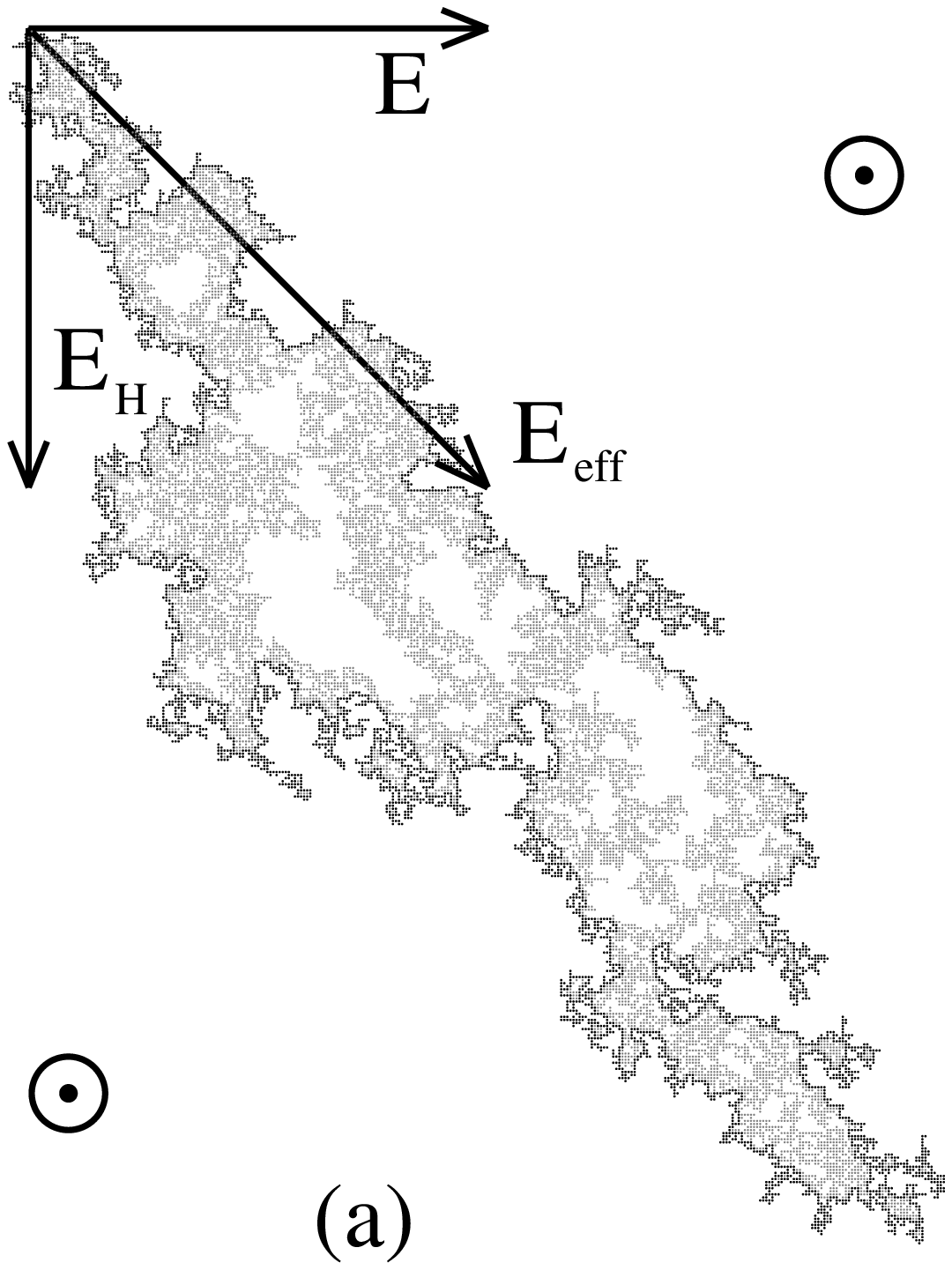,scale=0.4}
  \hfill\hfill \psfig{file=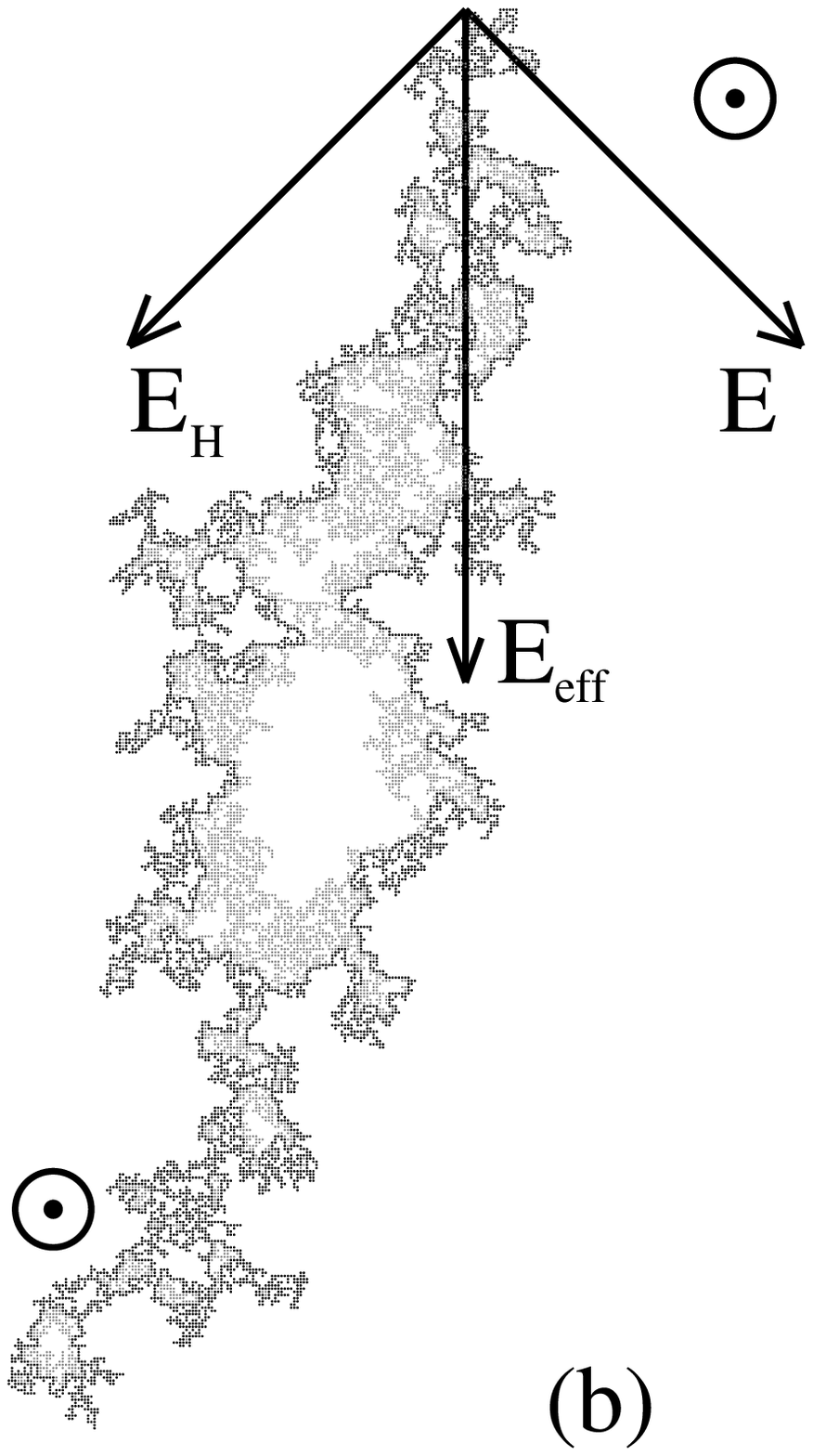,scale=0.4}\hfill}

\bigskip

\centerline{\hfill \psfig{file=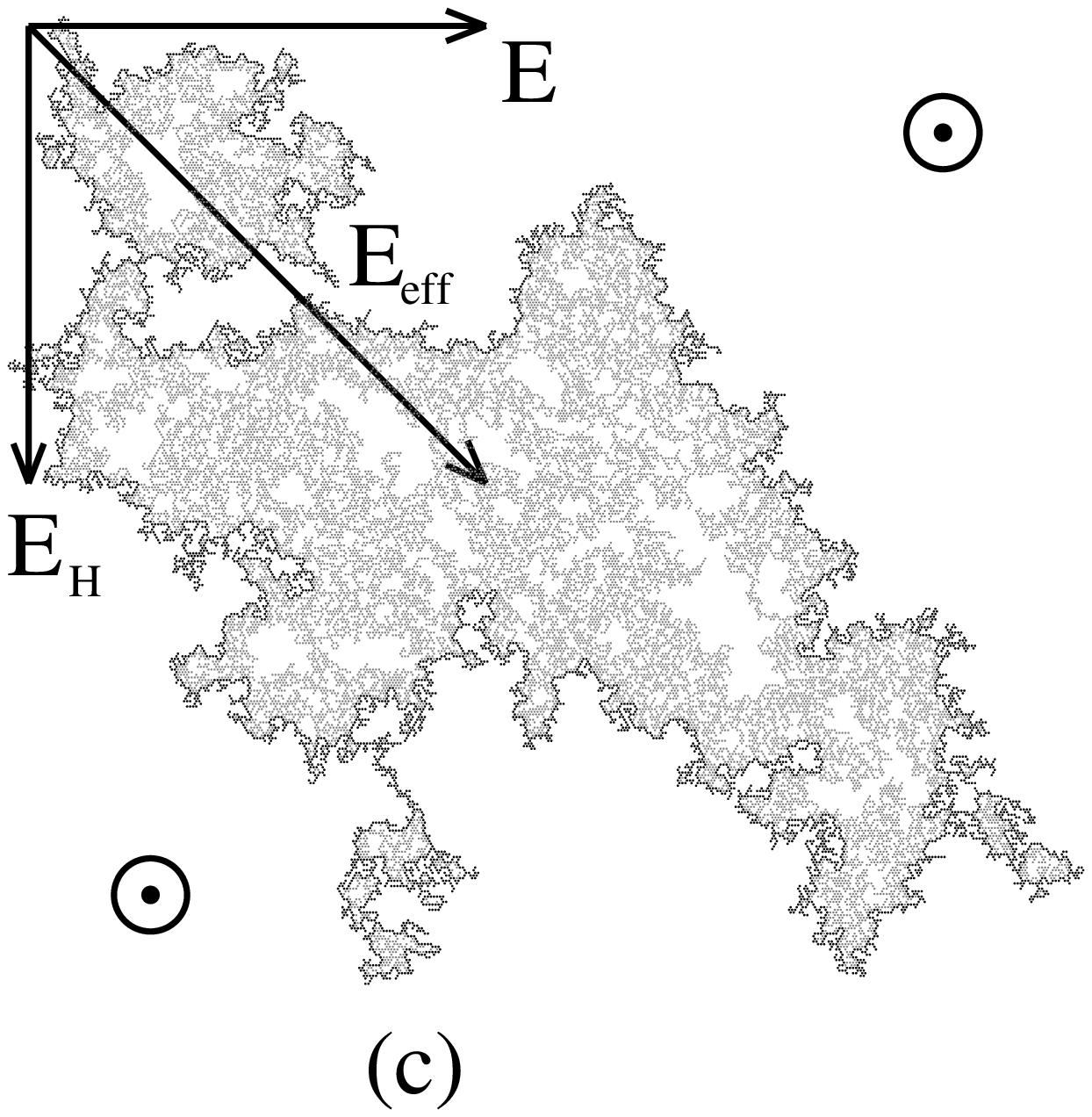,scale=0.4}
  \hfill\hfill \psfig{file=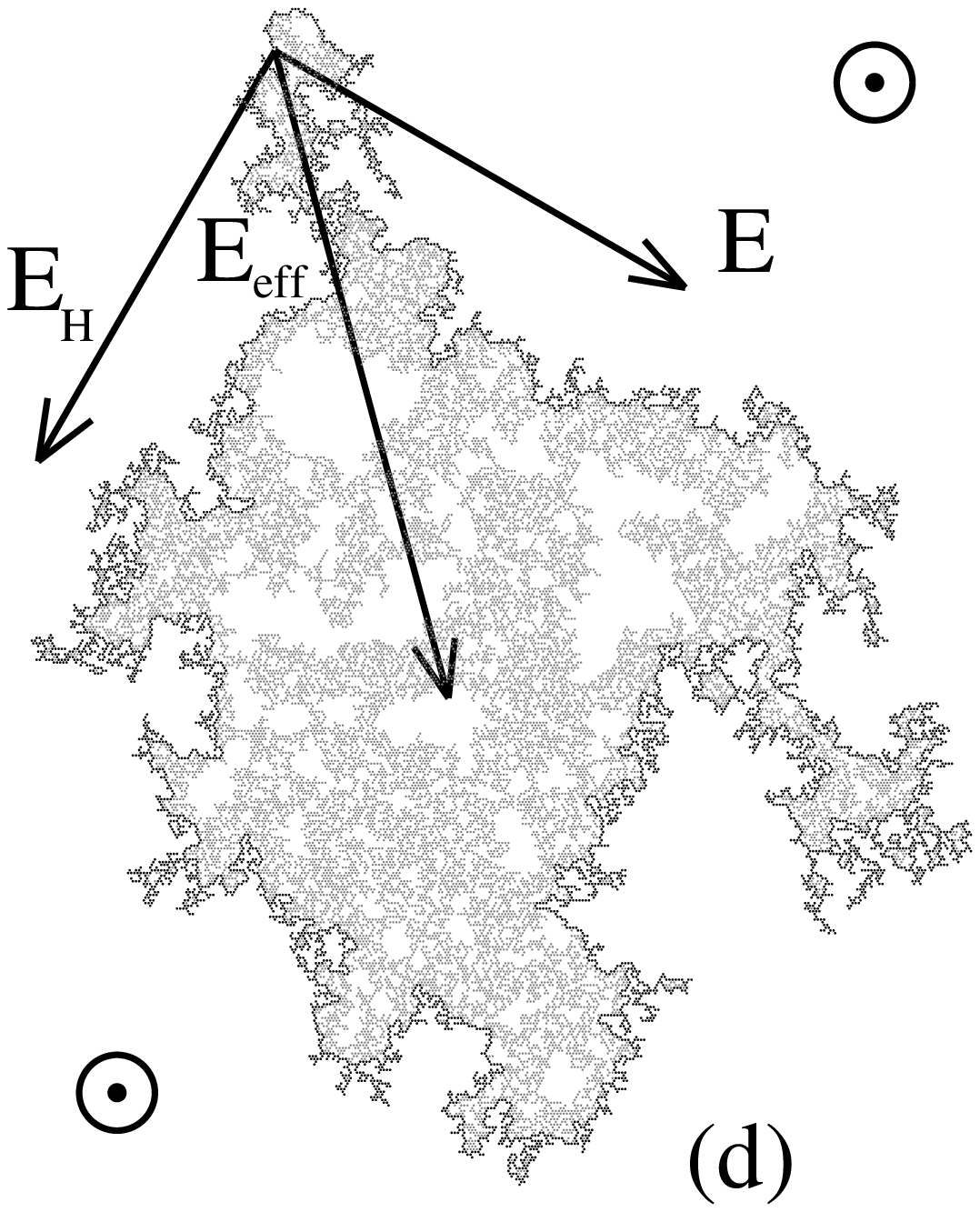,scale=0.4}\hfill}

\medskip
\caption{\label{cluster} Infinite spanning clusters at $p=p_c$ on the
  square ($a$, $b$) and triangular ($c$, $d$) lattices of size
  $L=2^8$.  For the square lattice, the $E$ field orientation is
  horizontal in ($a$) and it is diagonal in ($b$). For the triangular
  lattice, the $E$ field orientation is horizontal in ($c$) and it is
  semi-diagonal ($30^\circ$ with the horizontal) in ($d$). For all $E$
  field configurations, the $B$ field is directed into the plane of
  the lattice. The field lines are drawn from the origin of the
  clusters. $E_H$ is the Hall field normal to both $E$ and
  $B$. $E_{\sf eff}$ is the resultant field of $E$ and $E_H$. The gray
  dots represent the interior of the clusters and the black dots
  represent the cluster hull. The clusters are elongated along the
  effective field. }
\end{figure}

Typical spanning clusters at $p=p_c$ for different field
configurations on the square and triangular lattices of size $L=2^8$
are shown in Fig.\ref{cluster}. Spanning clusters generated on the
square lattice for the horizontal and diagonal $E$ field orientations
are represented in Fig.\ref{cluster}($a$) and in
Fig.\ref{cluster}($b$) respectively.  Fig.\ref{cluster}($c$) and
Fig.\ref{cluster}($d$) represent the spanning clusters on the
triangular lattice for the horizontal and semi-diagonal ($30^\circ$
with the horizontal) $E$ field orientations respectively. The $B$
field is always perpendicular to the $E$ field and directed into the
plane of the lattice as shown by encircled dots. The black dots
indicate the external boundary of the cluster and the gray dots
represent the interior of the cluster. It can be seen that the
clusters are highly rarefied, anisotropic and have chiral dangling
ends. Interestingly, the elongation of the clusters are not along the
applied $E$ field. This is due to the generation of the Hall field
$E_H$ perpendicular to the applied fields $E$ and $B$. As a result, an
effective field $E_{\sf eff}$ develops in the system and the
elongation of the clusters are along the effective field. Since the
direction of the effective field depends on the direction of the
applied $E$ field, the orientation of the cluster grown then depends
on the direction of the applied $E$ field. However, the clusters are
not merely DP clusters along the effective field. Because, the
presence of the rotational $B$ field gives rise to chiral dangling
ends at the boundary which has its own critical
behaviour\cite{santrahl}. It is important to notice that the clusters
are less anisotropic and more compact on the triangular lattice than
on the square lattice. This is due to the extra flexibility given in
the $B$ field on the triangular lattice. However, the clusters look
indifferent under different $E$ field orientations on a given lattice
except their direction of growth.

\subsection{Fractal dimension}
\begin{figure}
\centerline{\hfill \psfig{file=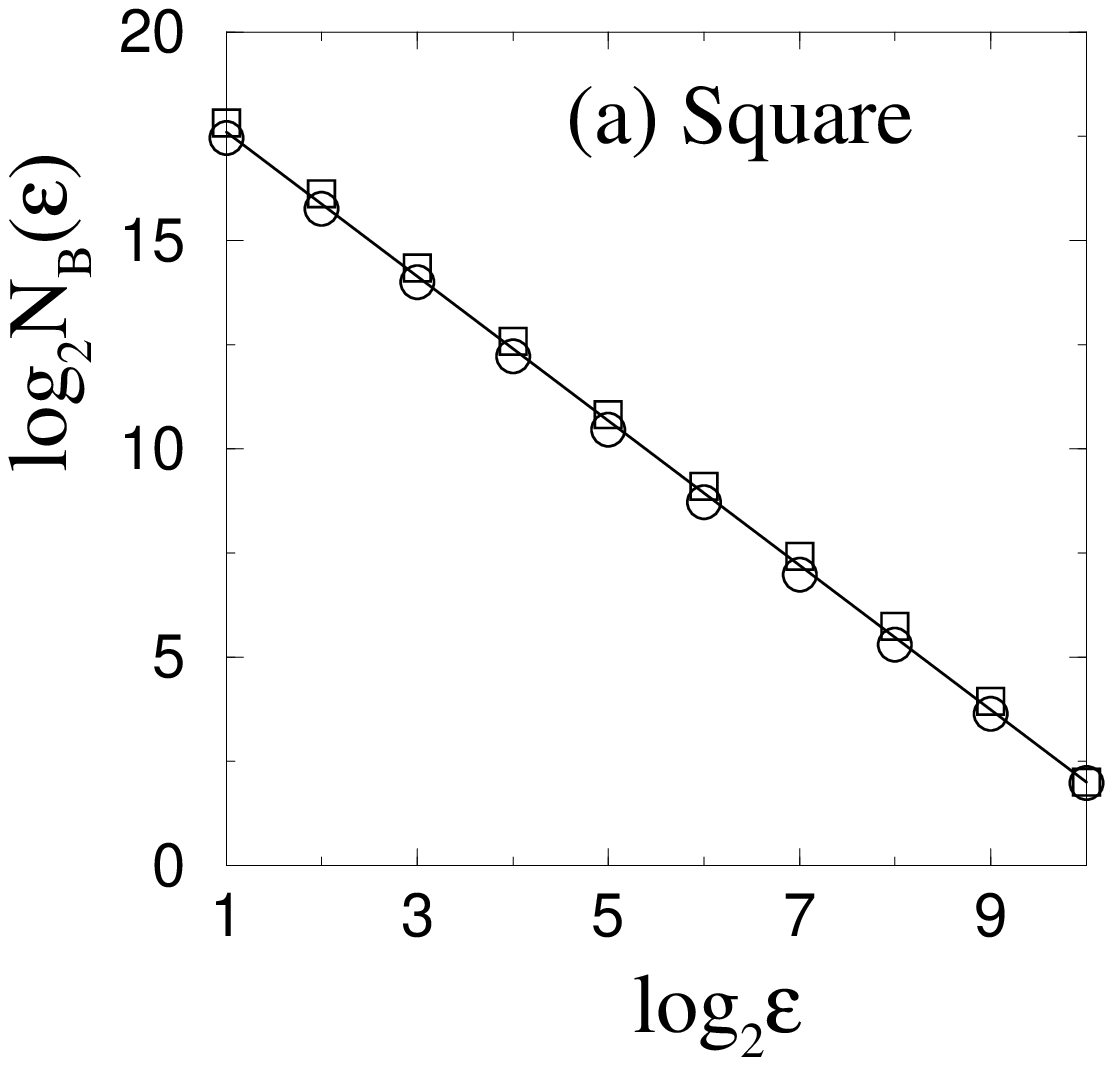,width=0.4\textwidth}
  \hfill\hfill \psfig{file=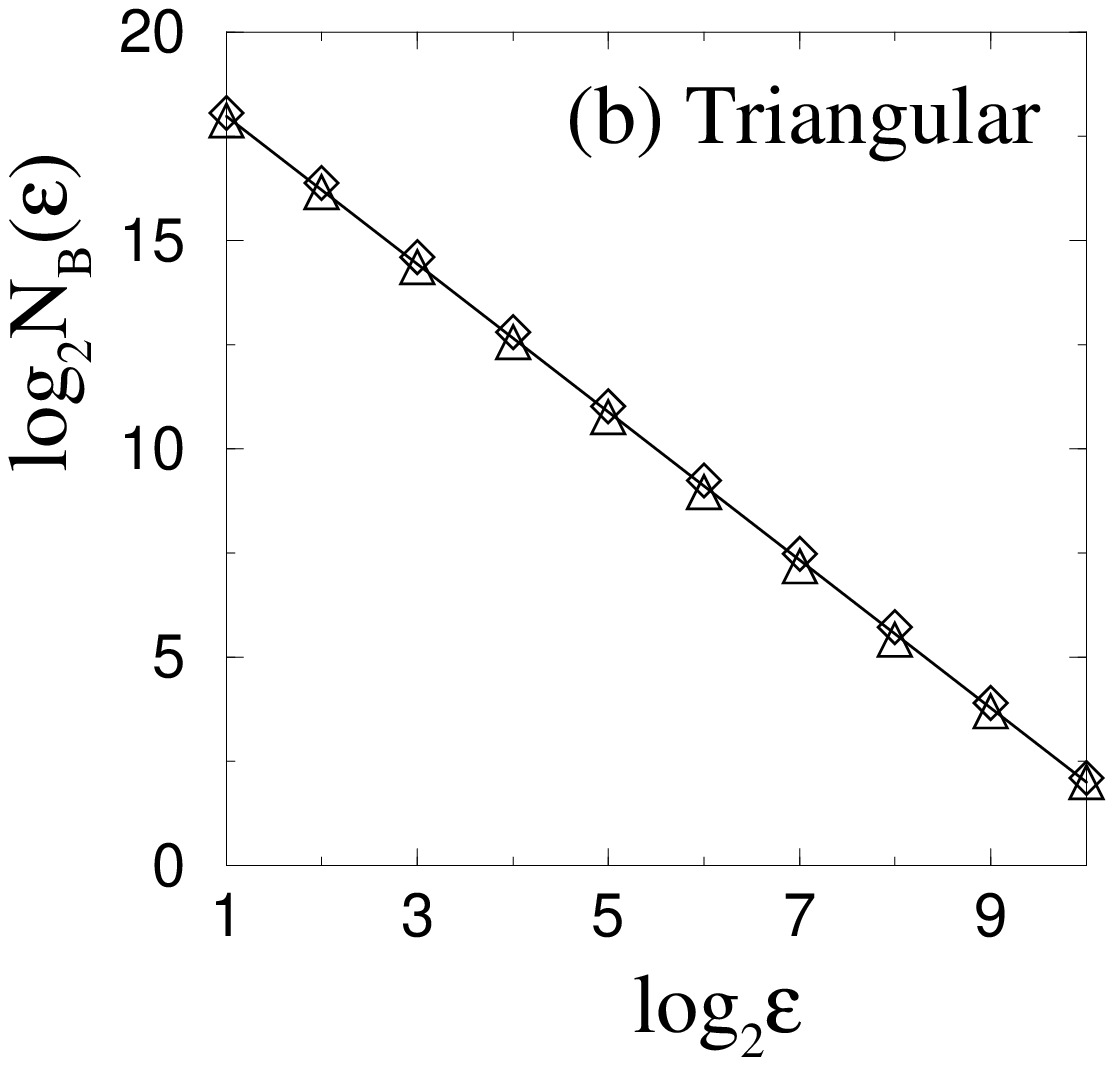,width=0.4\textwidth}\hfill}
\medskip
\caption{\label{fracd} Number of boxes $N_B(\epsilon)$ is plotted
  against the box size $\epsilon$ for the square ($a$) and triangular
  ($b$) lattices to determine the fractal dimension $d_f$ of the
  spanning clusters. Squares are for the horizontal $E$ and circles
  are for the diagonal $E$ field on the square lattice. Triangles are
  for the horizontal $E$ and diamonds are for the semi-diagonal $E$
  field on the triangular lattice. The solid lines are guide to
  eye. The values of $d_f$ are obtained as $d_f=1.732 \pm 0.006$ for
  diagonal $E$ field on the square lattice and $d_f=1.777\pm 0.005$
  for the semi-diagonal $E$ field on the triangular lattice. The value
  $d_f$ is then independent of the $E$ field orientation.}
\end{figure}

The fractal dimension $d_f$ of the infinite or spanning clusters are
determined by the box counting method. The number of boxes
$N_B(\epsilon)$ grows with the box size $\epsilon$ as $N_B(\epsilon)
\sim \epsilon^{d_f}$ where $d_f$ is the fractal dimension. In
Fig.\ref{fracd}, $N_B(\epsilon)$ is plotted against the box size
$\epsilon$ for the square ($a$) and triangular ($b$) lattices for all
$E$ field orientations. On the square lattice, the fractal dimensions
are found as $d_f= 1.733\pm 0.005$ for the horizontal $E$ field and
$d_f= 1.732\pm 0.006$ for the diagonal $E$ field. Similarly on the
triangular lattice, it is found that $d_f= 1.775\pm 0.004$ for the
horizontal $E$ field and $d_f= 1.777\pm 0.005$ for the semi-diagonal
$E$ field. The error is due to the least square fit. These estimates
are also confirmed by finite size (FS) analysis. In FS analysis, $d_f$
is determined employing $S_\infty \sim L^{d_f}$, where $S_\infty$ is
the mass of the spanning cluster. The FS estimates are $d_f=1.72\pm
0.02$ on the square lattice and $d_f = 1.81\pm 0.02$ on the triangular
lattice for the diagonal and semi-diagonal $E$ field respectively. It
is interesting to notice that the change in the direction of the $E$
field has no effect on the values of $d_f$ on both the
lattices. However, the values of $d_f$ are different on the square and
triangular lattices as already reported in Ref.\cite{sinha}. It was
already found that $d_f$ is higher on the triangular lattice than that
of on the square lattice consistent with the observation that the
spanning clusters are more compact on the triangular lattice than on
the square lattice. For a given $E$ field direction, the DSP model on
these two lattices differ only by an extra rotational direction due to
$B$ field on the triangular lattice than that on the square
lattice. The change in $d_f$ on these two lattices is then only due to
the different number of branching of the rotational constraint due to
the $B$ field in the presence of a directional field. The extra
flexibility of the rotational constraint allows the clusters to
penetrate more and more into itself on the triangular lattice than on
the square lattice. As a result, the infinite clusters are less
rarefied on the triangular lattice than on the square lattice. Since
the clusters grow along the effective field direction, the orientation
of the clusters is only changing with the orientation of the $E$ field
as the effective field direction is changing.  Note that, the value of
$d_f$ obtained here is not only different from other percolation
models but also smallest among that of OP ($91/48$\cite{nn}), DP
($1.765$\cite{hede}) and SP ($1.969$\cite{bose}). The spanning
clusters are then most rarefied in DSP. Since $d_f$ obtained in DSP is
different from that of DP clusters, it could be inferred that the DSP
clusters are just not DP clusters along the effective field. The DSP
clusters then should have its own critical behaviour at the
percolation threshold as it is already observed\cite{dsp}. It is now
important to check the effect of $E$ field direction on the other
critical exponents.

\subsection{Critical exponents and Scaling relations}
\begin{figure}
\centerline{\hfill \psfig{file=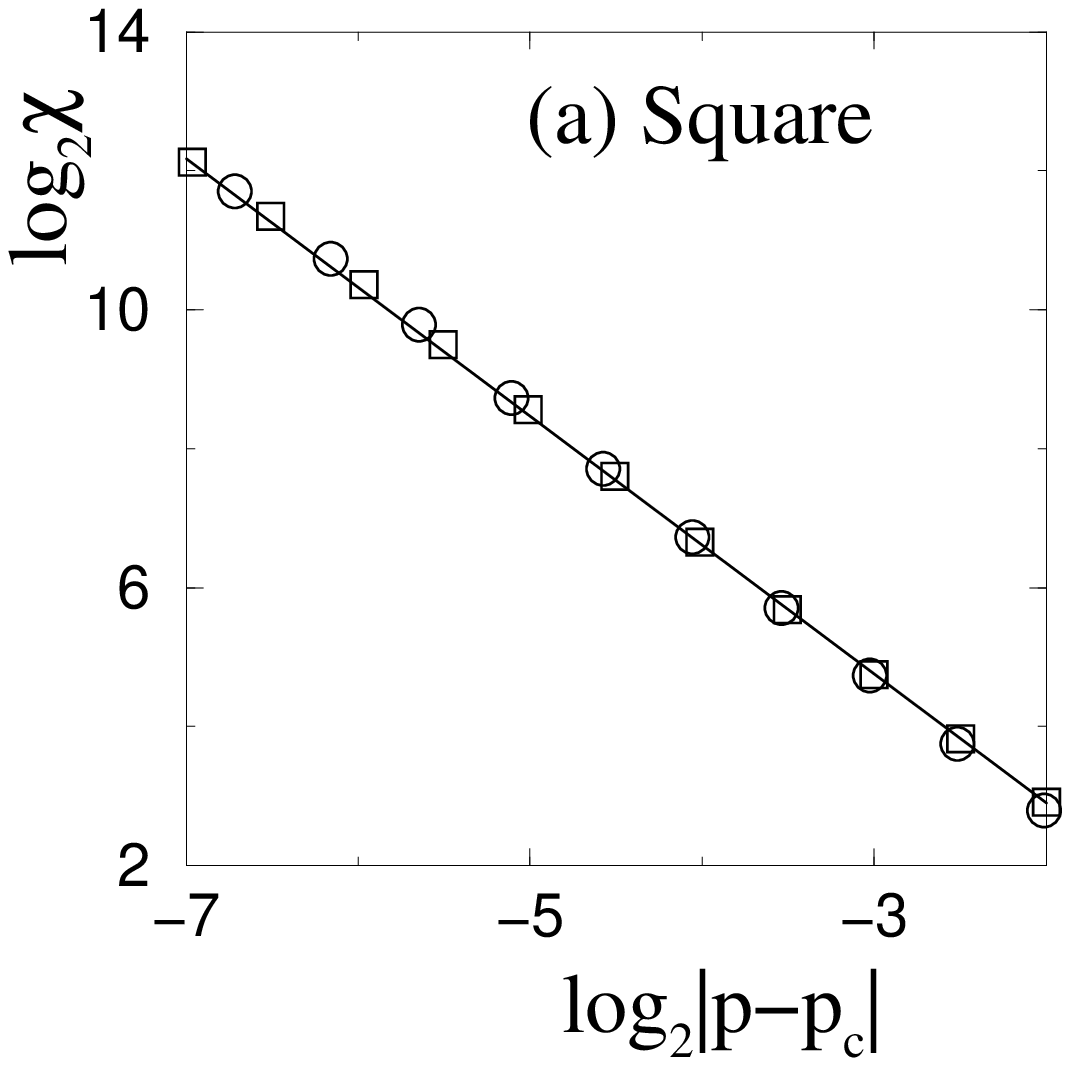,width=0.4\textwidth}
  \hfill\hfill \psfig{file=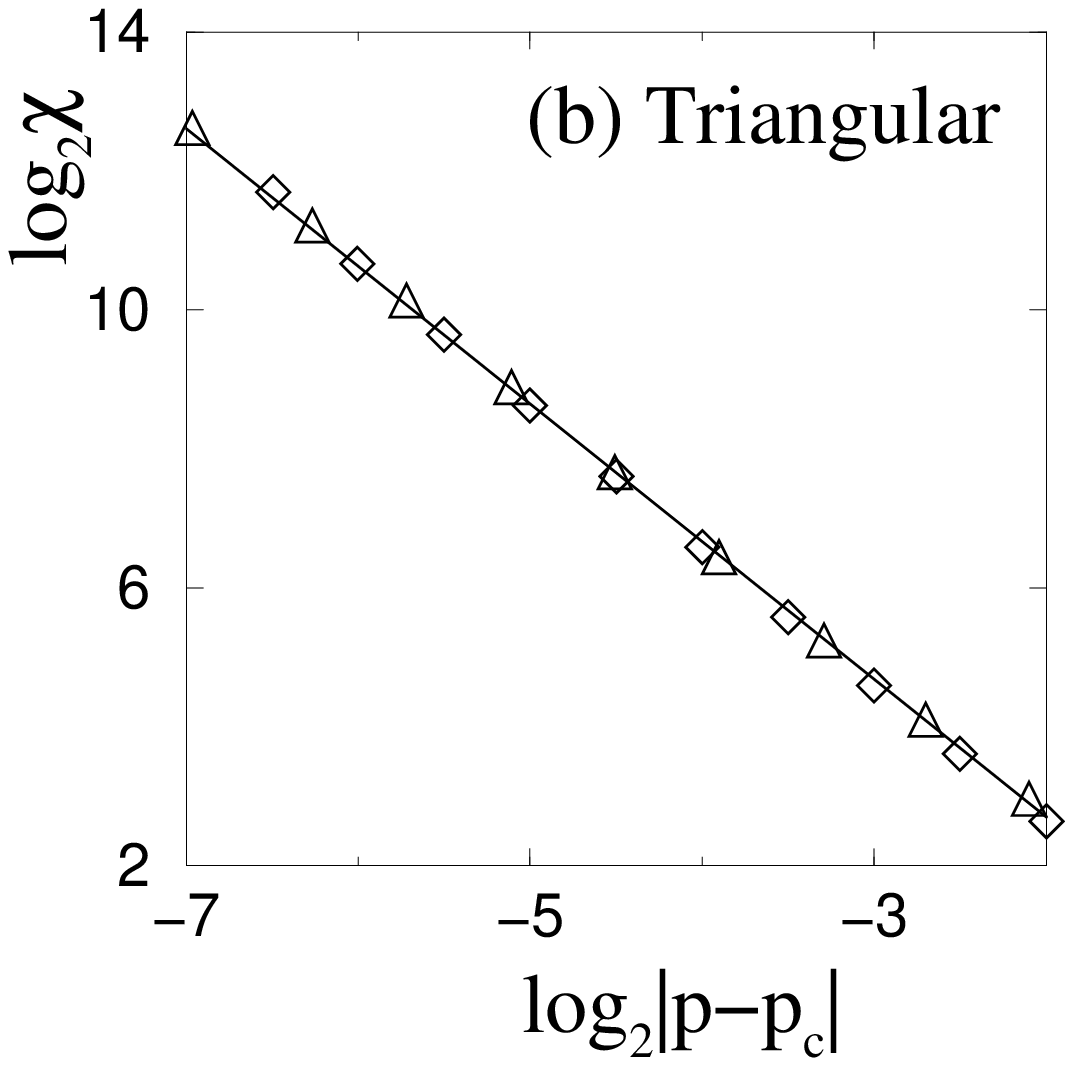,width=0.4\textwidth}\hfill }
\medskip  
\caption{\label{chi1} Plot of $\chi$ against $|p-p_c|$: ($a$) on the
   square and ($b$) on the triangular lattice for different $E$ field
   configurations. The same symbol set of the previous figure is
   used. The solid lines are fitted to the data corresponding to the
   diagonal $E$ field with slope $1.87\pm 0.01$ on the square
   lattice in ($a$) and to the semi-diagonal $E$ field with slope
   $2.00\pm 0.02$ on the triangular lattice in ($b$).}
\end{figure}

\begin{figure}
\centerline{\hfill \psfig{file=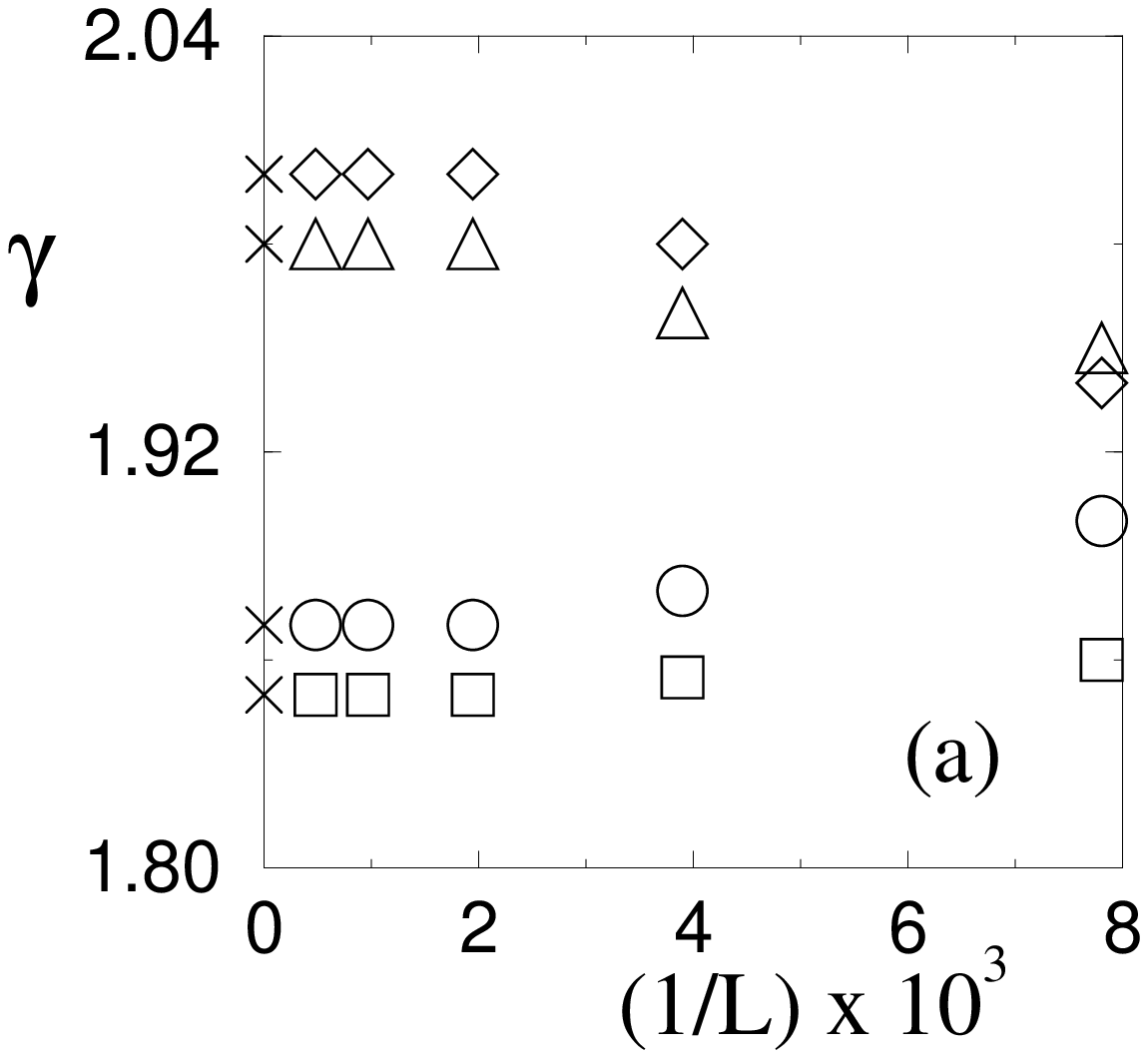,width=0.3\textwidth}
  \hfill\hfill \psfig{file=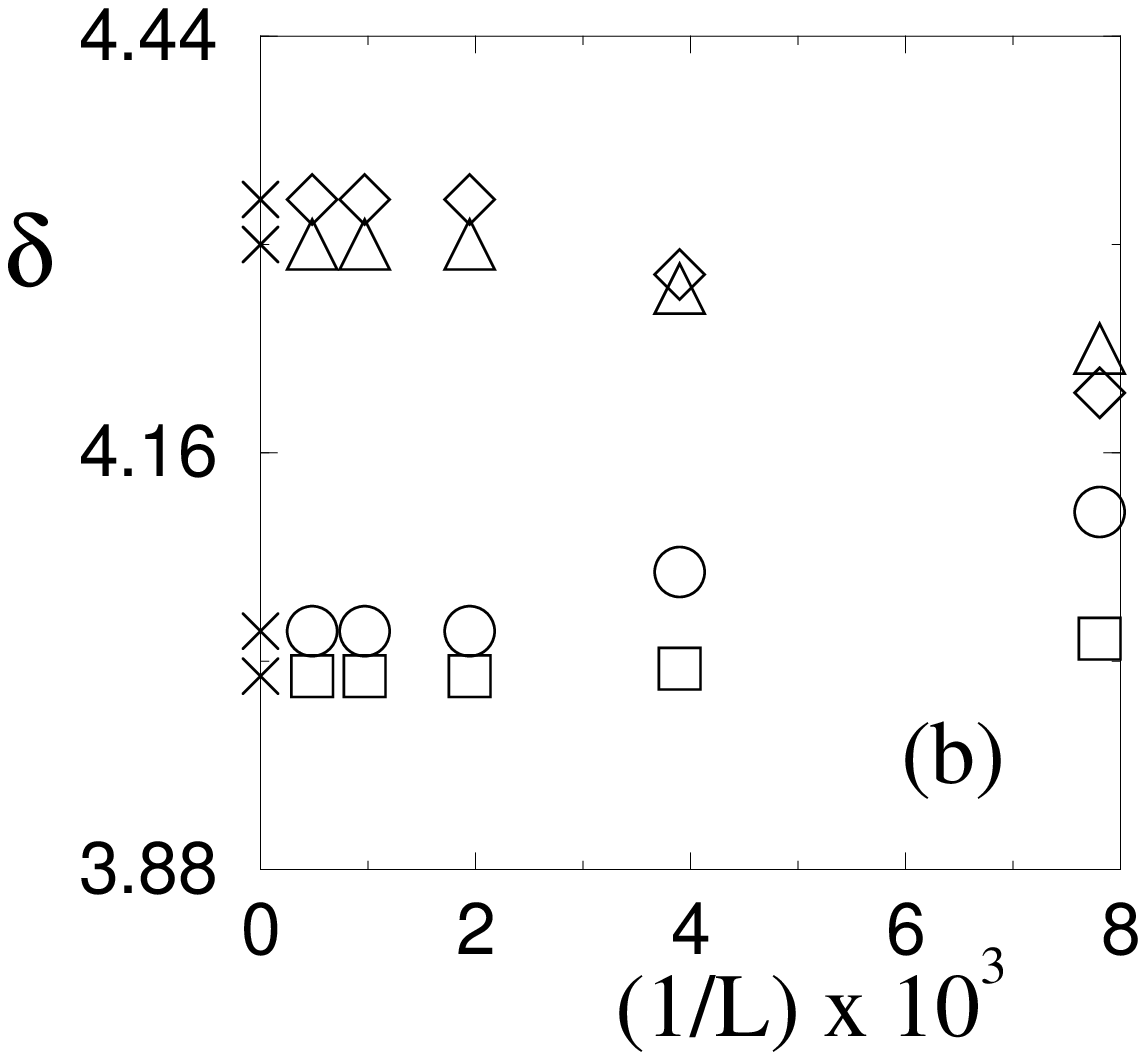,width=0.3\textwidth}
  \hfill\hfill \psfig{file=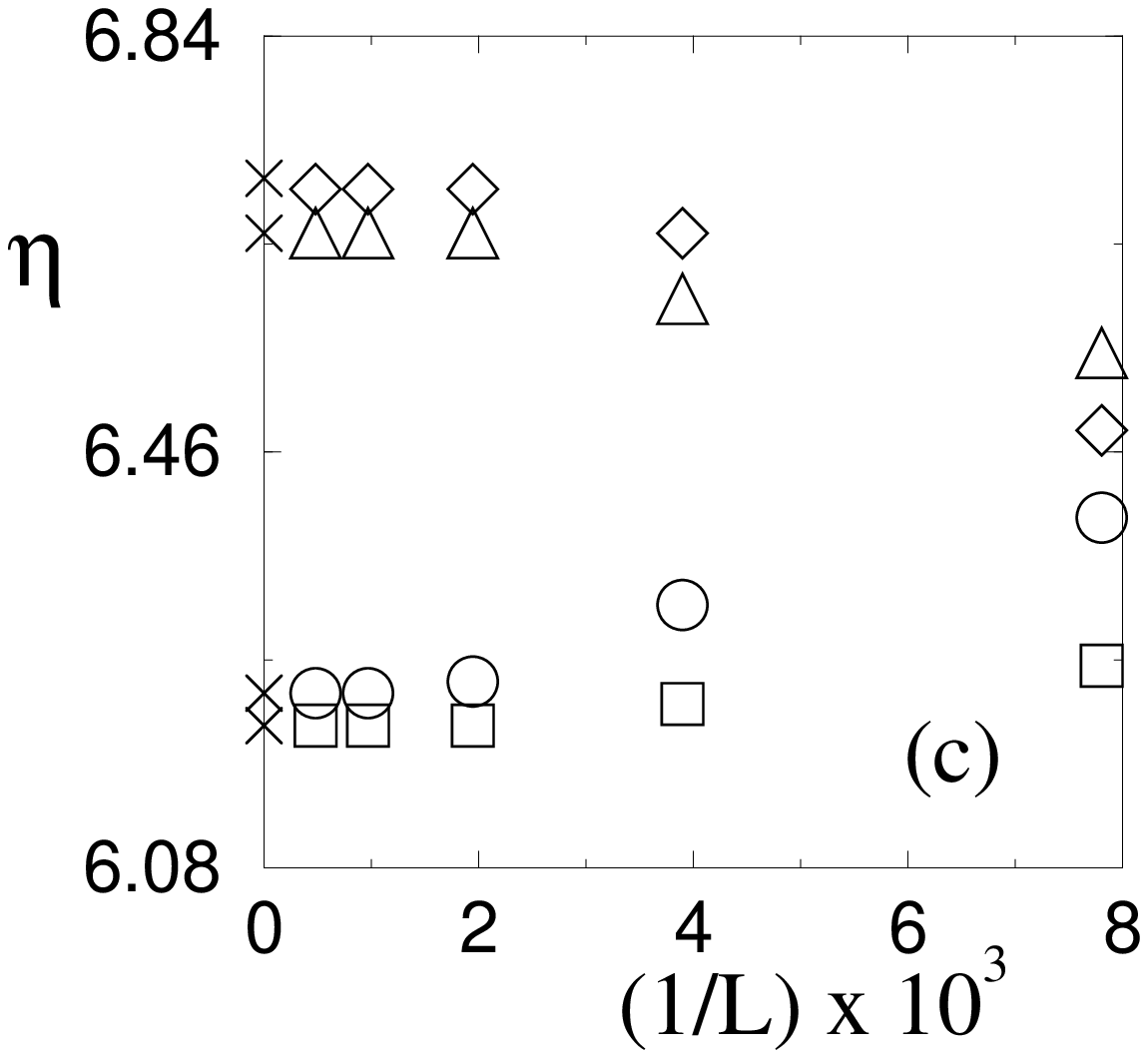,width=0.3\textwidth}\hfill }
\medskip
\caption{\label{gde} Plot of ($a$) $\gamma$, ($b$) $\delta$ and ($c$)
  $\eta$ against $1/L$ on the square ( $\Box$, $\bigcirc$) and
  triangular ($\triangle$, $\Diamond$) lattices for different $E$
  field configurations. The extrapolated values of the exponents are:
  $\gamma \approx 1.85$ and $1.87$, $\delta \approx 4.01$ and $4.04$
  and $\eta \approx 6.21$ and $6.24$ for horizontal and diagonal $E$
  field on the square lattice. On the triangular lattice, the values
  are: $\gamma \approx 1.98$ and $2.00$, $\delta \approx 4.30$ and
  $4.33$ and $\eta \approx 6.66$ and $6.71$ for the horizontal and
  semi-diagonal $E$ fields. The values of the critical exponents are
  within error bars for different $E$ field orientations.} 
\end{figure}

The values of some of the critical exponents are estimated and the
scaling relations among them are verified in this section. A full
scaling theory of DSP model is given in Ref.\cite{dsp}. The cluster
size distribution is defined as $P_s(p)=N_s/N_{tot}$ where $N_s$ is
the number of $s$-sited clusters in a total of $N_{tot}$ clusters
generated. Since the origin of a cluster is occupied with unit
probability, the scaling function form of the cluster size
distribution is assumed as
\begin{equation}
\label{scalef}
P_s(p)=s^{-\tau+1}{\sf f}[s^\sigma(p-p_c)]
\end{equation}
where $\tau$ and $\sigma$ are two exponents. The form of $P_s(p)$ has
already been verified and found appropriate for
DSP\cite{dsp,sinha}. In order to verify the scaling relations among
the critical exponents, the average cluster size, $\chi= \sum'_s
sP_s(p) \sim |p-p_c|^{-\gamma}$, and two other higher moments, $\chi_1
= \sum'_s s^2P_s(p) \sim |p-p_c|^{-\delta}$ and $\chi_2 = \sum'_s
s^3P_s(p) \sim |p-p_c|^{-\eta}$, of the cluster size distribution
$P_s(p)$ are measured generating finite clusters below $p_c$. $E$
field is applied diagonally on the square lattice and semi-diagonally
on the triangular lattice in the presence of a crossed $B$ field
directed into the plane of the lattice. For a system size $L=2^{11}$,
$\chi$ is plotted against $|p-p_c|$ for diagonal $E$ field on the
square lattice (circles) in Fig.\ref{chi1}($a$) and for semi-diagonal
$E$ field on the triangular lattice (diamond) in
Fig.\ref{chi1}($b$). Data corresponding to horizontal $E$ field on the
square (squares)\cite{dsp} and triangular (triangles)\cite{sinha}
lattices are also plotted in the respective figures for comparison
with the present data. It can be seen from Fig.\ref{chi1} that not
only the absolute magnitude of the average cluster size $\chi$ but
also the scaling of $\chi$ with $|p-p_c|$ remains independent of the
$E$ field orientation on both the lattices. The value of $\gamma$ is
found $1.87\pm 0.02$ for the the diagonal $E$ field whereas for the
horizontal $E$ field it was $1.85 \pm 0.02$ on the square
lattice. Thus, the value of the exponent $\gamma$ is within error bar
for both the $E$ field orientations on the square lattice. A similar
result is also obtained on the triangular lattice. It is found that
$\gamma=2.00\pm 0.02$ for semi-diagonal $E$ field whereas it was $1.98
\pm 0.02$ for the horizontal $E$ field on the triangular lattice.  The
errors quoted here include the least square fit error and the change
in $\gamma$ with $p_c\pm\Delta p_c$ for $\Delta
p_c=0.0005$. Similarly, estimates of the critical exponents $\delta$
and $\eta$ related to the higher moments of the cluster size
distributions are also obtained. The observations are listed in Table
\ref{cls_res}. It can be seen that the values of all the critical
exponents are very close and within error bars for different $E$ field
orientations on a given lattice. Finite size analysis of these
critical exponents $\gamma$, $\delta$ and $\eta$ have also been
made. The values of $\gamma$, $\delta$ and $\eta$ are determined on
different system size $L$. In Fig.\ref{gde}, the exponents are plotted
against the system size $L$. The extrapolated values of the exponents
in the $L\rightarrow \infty$ limit are marked by crosses. As expected,
the finite size estimates of $\gamma$, $\delta$ and $\eta$ are
converging to the Monte Carlo results on both the lattices as $L
\rightarrow \infty$. It is interesting to notice that the exponent
values are close and within error bars for different $E$ field
orientation on a given lattice although they are significantly
different (out of error bars) on the square and triangular
lattices. The exponents $\gamma$, $\delta$ and $\eta$ satisfy a
scaling relation $\eta = 2\delta - \gamma$. The scaling relation has
already been verified in Ref.\cite{dsp} and \cite{sinha} for
horizontal $E$ field and found holds true. The same scaling relation
has also been tested with the obtained values of the exponents for
diagonal/semi-diagonal orientations of $E$ field and it is satisfied
within error bars irrespective of $E$ field orientations on both the
lattices.

\begin{figure}
\centerline{\hfill \psfig{file=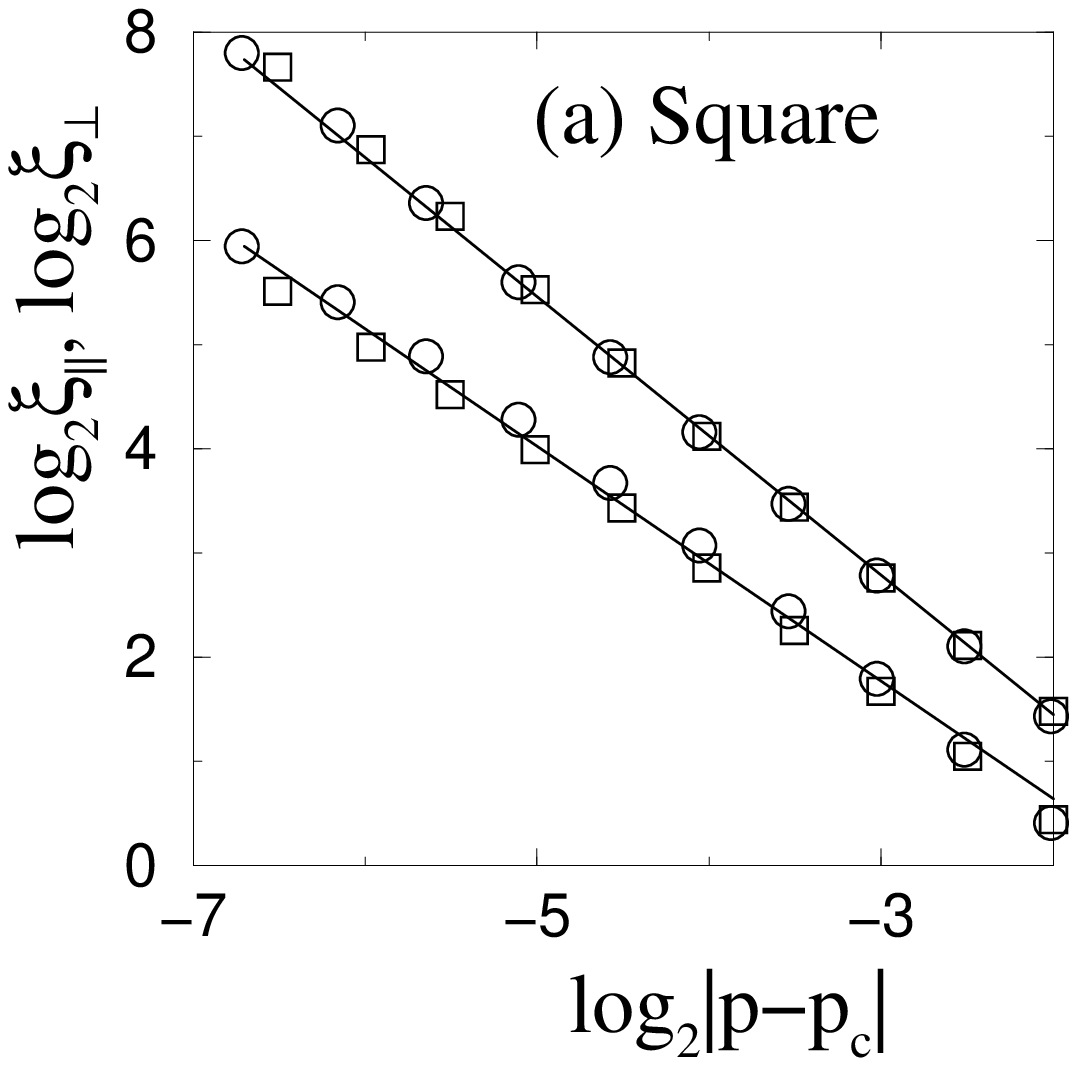,width=0.4\textwidth}
  \hfill\hfill \psfig{file=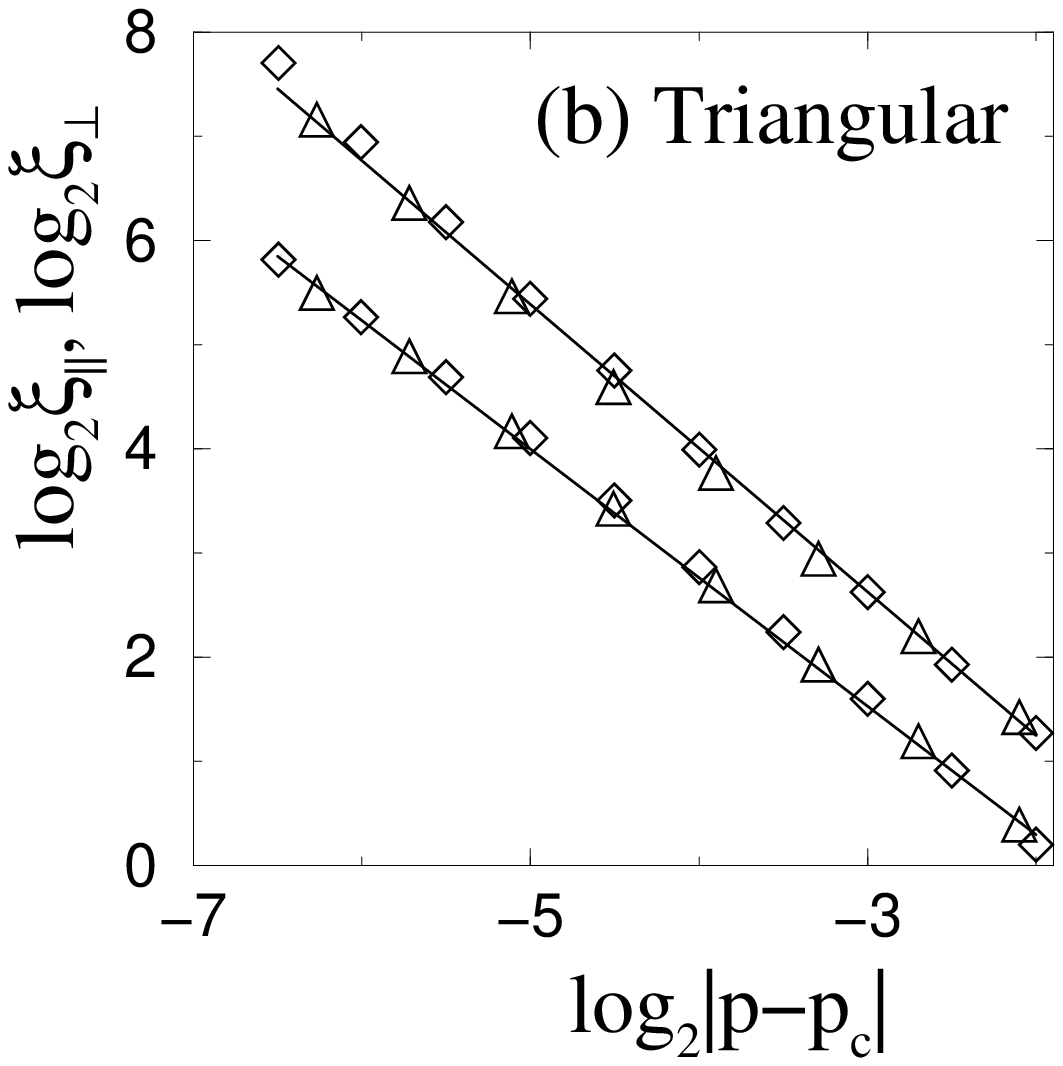,width=0.4\textwidth}\hfill}
\medskip
 \caption{\label{corrl} Plot of $\xi_\parallel$ and $\xi_\perp$
  against $|p-p_c|$: ($a$) on the square and ($b$) on the
  triangular lattice. The same symbol set of Fig.\ref{chi1} is used
  for different fields. The values of $\nu_\parallel$ and $\nu_\perp$
  are found with in error bars for different $E$ field orientations.}
\end{figure}

Since the clusters generated here are anisotropic, there are then two
connectivity lengths: $\xi_\parallel$ along the elongation of the
cluster and $\xi_\perp$ along the normal to the elongation. They are
defined as $\xi_\parallel^2 = 2\sum'_sR^2_\parallel(s)
sP_s(p)/\sum'_ssP_s(p)$ and $\xi_\perp^2 = 2\sum'_sR^2_\perp(s)
sP_s(p)/\sum'_ssP_s(p)$ where $R_\parallel(s)$ and $R_\perp(s)$ are
the two radii of gyration with respect to two principal axes.  The
connectivity lengths, $\xi_\parallel \sim |p-p_c|^{-\nu_\parallel}$
and $\xi_\perp \sim |p-p_c|^{-\nu_\perp}$ diverge with two different
critical exponents $\nu_\parallel$ and $\nu_\perp$ as $p\rightarrow
p_c$. The scaling behaviour of $\xi_\parallel$ and $\xi_\perp$ are
studied as function of $p$ below $p_c$ for diagonal $E$ field on the
square lattice and for semi-diagonal $E$ field on the triangular
lattice. For lattice size $L=2^{11}$, $\xi_\parallel$ and $\xi_\perp$
are plotted against $|p-p_c|$ in Fig.\ref{corrl}($a$) for the square
lattice and in Fig.\ref{corrl}($b$) for the triangular lattice for
different $E$ field orientations. The exponents $\nu_\parallel$ and
$\nu_\perp$ are obtained as $1.34\pm 0.02$ and $1.13\pm 0.02$
respectively for diagonal $E$ field on the square lattice
(circles). On the triangular lattice, they are found as $\nu_\parallel
= 1.37\pm 0.02$ and $\nu_\perp = 1.24\pm 0.02$ for the semi-diagonal
$E$ field (diamonds). Data for the horizontal $E$ field are indicated
by squares for the square lattice and by triangles for the triangular
lattice. The values of $\nu_\parallel$ and $\nu_\perp$ are found with
in error bars for different $E$ field orientations on both the
lattices, see Table \ref{cls_res}. The errors quoted here are the
least square fit errors. Finite size estimates of these exponents are
also made and it is found that they are consistent with that of the MC
results. The exponent $\nu_\parallel$ is almost same on the square and
triangular lattices whereas $\nu_\perp$ on the triangular lattice is
little higher than that of the square lattice value. $\Delta\nu =
\nu_\parallel -\nu_\perp$ is $\approx 0.21$ for DSP on the square
lattice whereas it is $\approx 0.64$ in DP\cite{nudp}. The DSP
clusters are then less anisotropic than that of DP clusters. This
observation remains unchanged even if the direction of the effective
field, along which the clusters grow, is changed. It is also noticed
that the critical behaviour of the connectivity length $\xi_\parallel$
along the elongation of the cluster is similar to that of the
connectivity length $\xi$ of ordinary percolation. The value of
$\nu_\parallel$ is almost equal to $\nu(OP) = 4/3$\cite{nn}. This is
in good agreement with a recent study of magnetoresistance of a
three-component composites of metallic film by Barabash {\em et
al}\cite{cmat}. It is found that the hyperscaling relations $2\delta -
3\gamma = (d-1)\nu_\perp +\nu_\parallel$ and $(d-d_f)\nu_\perp =
\delta-2\gamma$ are satisfied on the square lattice within error bars
whereas they are satisfied marginally on the triangular lattice. It is
already known that the hyperscaling is violated in DP\cite{hp}.

The results obtained for different $E$ field orientations on the
square and triangular lattices are summarized in Table
\ref{cls_res}. Comparing the data given in Table \ref{cls_res} it can
be concluded that the values of the critical exponents and the scaling
relations among them are independent of $E$ field orientation or the
number of components of $E$ field irrespective of lattice
structure. The direction of $E$ field is only able to change the
orientation of the clusters and not the critical properties. However,
it has already been observed that the values of the critical exponents
obtained in DSP model are not only different from that of the other
percolation models but also are significantly different on the square
and triangular lattices in $2D$. It is already reported in
Ref.\cite{sinha} as a breakdown of universality between the square and
triangular lattices. Microscopically, the DSP model on the square and
triangular lattices differs only by an extra rotational direction due
to $B$ field on the triangular lattice than that on the square lattice
for a given $E$ orientation. The values of the critical exponents then
depend on the number of branching due to the rotational field $B$ but
do not depend on the number of branching due to the directional field
$E$. By reducing the number of branching of $B$ field from three to
two on the triangular lattice and keeping $E$ field horizontal, an
artificial model has been created and the critical exponents are
re-estimated. The values of the critical exponents for this particular
field combination are obtained as: $d_f=1.730 \pm0.005$, $\gamma=1.87
\pm 0.01$, $\delta=4.08 \pm 0.04$ and $\eta=6.34 \pm
0.08$. Remarkably, the values of $d_f$ and other critical exponents
are found within error bars of that of the square lattice with
horizontal $E$ field and the universality holds. It should be
mentioned here that no breakdown of universality was observed in
SP\cite{bose}, percolation under only rotational constraint, between
the square and triangular lattices. In DSP model, the presence of the
directional constraint $E$ on top of the rotational constraint $B$ may
increase the number of spiral lattice trees. Since the spiral lattice
trees\cite{santrala} as well as the spiral self avoiding
walks\cite{saw} have different scaling behaviour on the square and
triangular lattices, the higher number of tree like structures in the
clusters generated can induced a non-trivial effect on the critical
properties of DSP at the percolation threshold and consequently the
discrepancy in the exponent values on the two lattices could occur.

Though there is a breakdown of universality in the cluster property of
the DSP model on the square and triangular lattices, it was already
observed that the universality holds true for their hull (external
perimeter) properties between the square and triangular
lattices\cite{hull}. The hull properties are also verified here by
changing the applied $E$ field direction and it is found that the hull
properties are independent of the $E$ field orientations.

\section{DSP Model under Variable Field Intensity}
So far, the DSP model has been investigated under constant external
field intensities. It is also useful to understand the properties of
disordered system as function of the field intensity. The intensities
$E$ and $B$ of the external fields can vary here from $0$ to $1$. It
may be noted that for certain extreme values of the intensities $E$
and $B$, the percolation model has four different varieties, DSP, DP,
SP and OP. DSP corresponds to $E=1$, $B=1$; DP corresponds to $E=1$,
$B=0$; SP corresponds to $E=0$, $B=1$ and OP corresponds to $E=0$,
$B=0$. Thus, a complete phase diagram can also be obtained by studying
the DSP model under variable field intensities between one and
zero. If the field intensities are changed continuously from ($1,1$)
to ($0,0$) keeping $E=B$, it is expected to have a phase change from
DSP to OP. Similarly, a phase change from DSP to SP can be obtained by
changing $E$ from $1$ to $0$ keeping $B=1$ and phase change from DSP
to DP can be obtained by changing $B$ from $1$ to $0$ keeping
$E=1$. In order to study the effect of intensities of the external
applied fields, one needs to generate percolation clusters under
variable field intensities and characterize their geometrical
properties at the percolation threshold. The intensity dependent
percolation cluster could be generated by selecting nearest neighbours
of an occupied site with appropriate probabilities, function of field
intensities. To demonstrate the model, a square lattice of size $L$ is
considered. The $E$ field is applied diagonally and the $B$ field is
applied perpendicular to the $E$ field and directed into the plane of
the lattice as shown in Fig.\ref{demo}(b). However, the results of
this particular field configuration should be valid for all other
field configurations. In order to occupy a site in the presence of
external fields, a list of eligible sites for occupation is prepared
first. The field intensities play role in the preparation of the list
of eligible sites. If the field intensities are either $E=1$ or $B=1$
or both $E=B=1$, the sites only in the favourable directions, defined
by the directional and rotational constraints, are eligible for
occupation. Sites in other directions are not considered for
occupation in all three constrained percolation models DSP, DP and
SP. The sites in the favourable directions of the fields are selected
with probabilities proportional to the field intensities. If the field
intensities are less than one, the sites in unfavourable directions of
the fields will also be eligible for occupation due to scattering. A
scattering field $S$ is then introduced in the model. Sites in
unfavourable directions are selected for occupation with probability
$S$ and it is defined as
\begin{equation}
S = (1-E)\times(1-B).
\label{ebselec}
\end{equation}
Notice that $S=0$ in all three cases: $E=1$, $B=1$ and $E=B=1$. That
means there is no scattering and consequently no sites in unfavourable
directions are eligible for occupation as in the case of DP, SP or
DSP. On the other hand, when $E$ and $B$ both are equal to zero then
$S=1$. There will be then no favourable direction. All empty nearest
neighbours of an occupied site on the square lattice are selected with
unit probability for occupation. The situation corresponds to
OP. Thus, a change of phase from DSP to OP, DP or SP will be possible
by changing $E$ and $B$ continuously. As soon as the list of eligible
sites is prepared they are then occupied with percolation probability
$p$. Below, the critical properties of DSP clusters are studied
varying intensity of $E$ and $B$ fields. For each value of $E$ and
$B$, the percolation threshold $p_c$ is determined. Data are averaged
over $10^5$ clusters for each field intensity.

\section{Effect of field intensity}

\begin{figure}
\centerline{\psfig{file=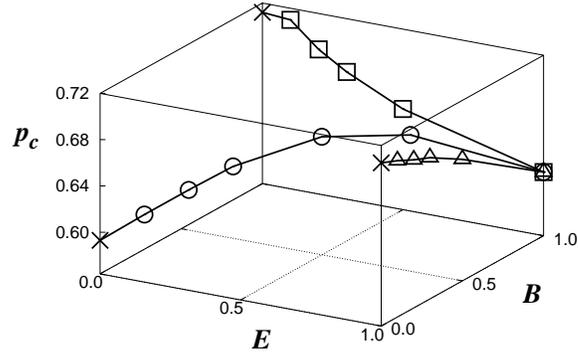,width=0.5\textwidth}}
\medskip
\caption{\label{pc_int} Plot of percolation threshold $p_c$ against
field intensities $E$ and $B$ for a square lattice of size
$L=1024$. For appropriate values of $E$ and $B$, $p_c$s correspond to
that of respective percolation models.}
\end{figure}

The field intensities are changed in three different ways. First, $E$
and $B$ are changed from $(1,1)$ to $(0,0)$ keeping $E=B$ and
accordingly DSP is expected to change over to OP. The corresponding
data will be represented by circles. Second, keeping $B=1$, $E$ is
changed from $1$ to $0$, then DSP should change to SP. These data are
represented by squares. Third, keeping $E=1$, $B$ is changed from $1$
to $0$, thus DP will be reached from DSP. Triangles represent the
corresponding data. Percolation threshold $p_c$, at which a spanning
cluster appears for the first time in the system, has been calculated
as function of the field intensities $E$ and $B$. Following the same
procedure of section $3$, $p_c$s are identified as the percolation
probability $p$ corresponding to the maximum slope of the plot of
spanning probability $P_{sp}$ against $p$. In Fig.\ref{pc_int}, the
values of $p_c$s are plotted against $E$ and $B$ for a square lattice
of size $L=1024$. Each $p_c$ has been determined generating $10^5$
spanning clusters. It can be seen that as $E$ and $B$ goes to zero
$p_c$ coincides with that of OP. Similarly, $p_c$ for DP and SP are
also obtained for $E=1, B=0$ and $E=0, B=1$ respectively.

\begin{figure}
\centerline{  \psfig{file=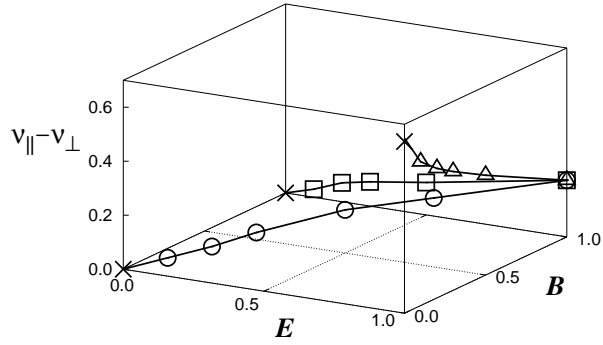,width=0.5\textwidth} }
\medskip
\caption{\label{dnu_int} Plot of $\Delta\nu = \nu_\parallel-\nu_\perp$
against field intensities $E$ and $B$. $\Delta\nu$ continuously goes
to zero as the field intensities goes to $E=B=0$ for OP and $B=1$,
$E=0$ for SP respectively. $\Delta\nu$ is also approaching $0.64$
corresponding to DP for $E=1$ and $B=0$.}
\end{figure}

The percolation clusters are isotropic for OP and SP whereas they are
anisotropic in the case of DP and DSP. For anisotropic clusters, there
are two connectivity lengths $\xi_\perp$ and $\xi_\parallel$ and for
isotropic clusters there is only one connectivity length. Since
$\xi_\perp$ and $\xi_\parallel$ diverge with two critical exponents
$\nu_\perp$ and $\nu_\parallel$ respectively as $p\rightarrow p_c$,
the ratio of two connectivity lengths $\xi_\parallel/\xi_\perp$ should
diverge as $|p-p_c|^{-\Delta\nu}$ where $\Delta\nu = \nu_\parallel -
\nu_\perp$. For isotropic clusters, $\nu_\parallel$ is equal to
$\nu_\perp$ since the clusters grow equally in all directions. Thus,
$\Delta\nu$ is zero for OP and SP clusters and $\Delta\nu$ is finite
for anisotropic clusters of DP and DSP. In DSP, $\Delta\nu$ is
approximately $0.21$\cite{dsp} whereas in DP, it is approximately
$0.64$\cite{nudp}. The values of the connectivity exponents
$\nu_\perp$ and $\nu_\parallel$ are determined for different values of
$E$ and $B$. In Fig.\ref{dnu_int}, $\Delta\nu$ is plotted against $E$
and $B$. As the field intensities are taken to $E=0$ and $B=0$ or
$E=0$ and $B=1$ corresponding to OP and SP models the value of $\Delta
\nu$ continuously goes to zero. As the field intensities are taken to
$E=1$ and $B=0$ corresponding to the DP model, the value of $\Delta
\nu$ approaches to $0.64$ as it is expected for DP. Not only the
difference but also the absolute value of the connectivity exponent
matches with the respective percolation models. Thus, there is a
smooth crossover from DSP to OP, DP and SP at the appropriate values
of $E$ and $B$ as they are changed continuously.

\begin{figure}
\centerline{  \psfig{file=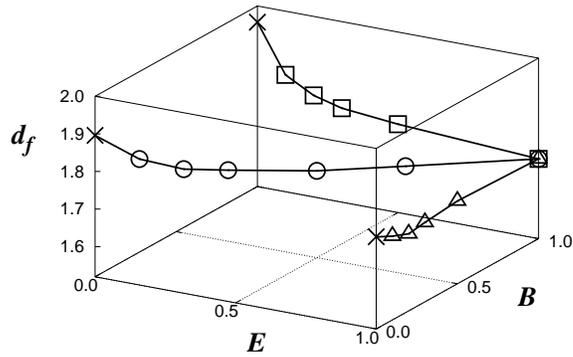,width=0.5\textwidth} }
\medskip
\caption{\label{df_int} Plot of fractal dimension $d_f$ of the
spanning clusters against field intensities $E$ and $B$. As $E$ and
$B$ change continuously, the value of $d_f$ approaches to that of the
respective percolation models at the appropriate values of $E$ and
$B$. }
\end{figure}

\begin{figure}
\centerline{  \psfig{file=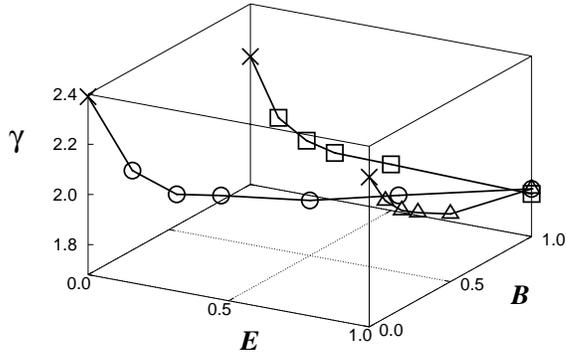,width=0.5\textwidth} }
\medskip
\caption{\label{gm_int} Plot of average cluster size exponent $\gamma$
against field intensities $E$ and $B$. $\gamma$ changes continuously
to that of the respective percolation models at the appropriate values
of $E$ and $B$.}
\end{figure}

In order to determine other critical properties, fractal dimension
$d_f$ of the spanning clusters and average cluster size exponent
$\gamma$ are also determined. $d_f$ is plotted in Fig.\ref{df_int} and
$\gamma$ is plotted in Fig.\ref{gm_int} against $E$ and $B$. The value
of $d_f$ in different percolation model are: $d_f(OP) = 91/48 \approx
1.896$\cite{nn}, $d_f(DP) \approx 1.765$\cite{hede}, $d_f(SP) \approx
1.957$\cite{bose} and $d_f(DSP) \approx 1.733$\cite{dsp}. The values
of $d_f$ for OP, DP and SP are marked by crosses in
Fig.\ref{df_int}. It can be seen that the value of $d_f$ changes
continuously and approaches to the expected values of $d_f$ of the
respective percolation models as the intensities of $E$ and $B$ are
changed appropriately. In Fig.\ref{gm_int}, the values of $\gamma$ of
different percolation models are also indicated by crosses,
$\gamma(OP)= 43/18 \approx 2.389$\cite{nn}, $\gamma(DP) \approx
2.2772$\cite{hede}, $\gamma(SP) \approx 2.19$\cite{bose} and
$\gamma(DSP)= 1.85$\cite{dsp}. Similar to the fractal dimension, the
value of the critical exponent $\gamma$ is also changing continuously
to that of the other percolation models as the intensities of $E$ and
$B$ are changed appropriately. However, for most of the $E$ and $B$
values, the values of $d_f$ and other critical exponents are found
close to that of DSP. It could also be noticed that in the presence of
small field intensities of $E$ or $B$, the universality class of OP is
changed. Similarly for DP and SP, a small intensity of the other field
is able to change the universality class of the models to that of
DSP. This is a new and important observation.

\begin{figure}
\centerline{  \psfig{file=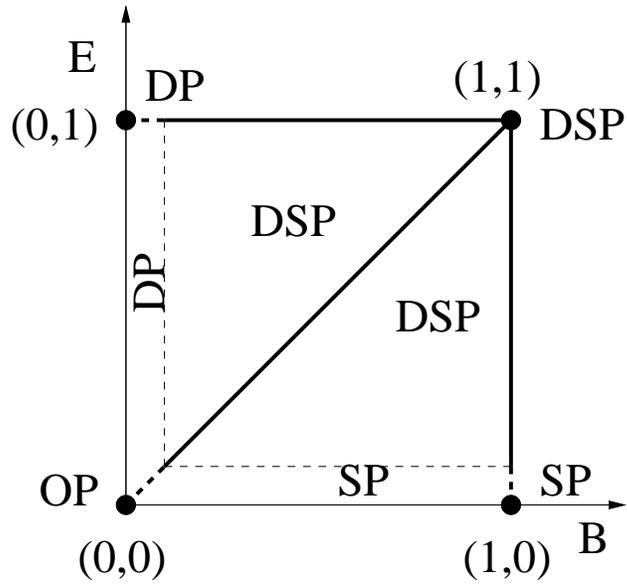,width=0.5\textwidth} }
\medskip
\caption{\label{pspc} Phase diagram of the percolation models under
  crossed external bias fields. Different percolation models are
  represented by black circles. DSP changes continuously to OP, DP or
  SP for the field intensities of $E$ and $B$ are changed to the
  appropriate values. }
\end{figure}

A phase diagram for percolation models under external bias fields now
can be plotted in the space of directional ($E$) and rotational ($B$)
bias fields. Different percolation models are identified on the $E-B$
space by black circles in Fig.\ref{pspc}. Each point in this space
corresponds to a second order phase transition point. The thick lines
from left lower to right upper diagonal and parallel to the $E$ and
$B$ axes of the phase diagram represent lines of second order phase
transition points. The two ends of the diagonal line correspond to OP
and DSP. The line parallel to $E$ connects SP and DSP whereas the line
parallel to $B$ connects DP and DSP. The dotted lines define the
regions of DP and SP as indicated. DSP model then can be considered as
the most generalized model of percolation under external bias fields
whose different limiting situations correspond to other percolation
models like OP, DP or SP.

\section{Summary}

The directed spiral percolation model has been studied varying the
applied field directions and their intensities on the square and
triangular lattices in $2D$. The critical properties of the cluster
related quantities in this model were already found very different
from the other percolation models like OP, DP and SP and accordingly
DSP belongs to a new universality class. It is found here that the
critical properties as well as the universality class of the DSP
clusters remain invariant on the directions of the applied $E$ and $B$
fields. Change of $E$ field direction corresponds to the change in the
number of components of the $E$ field on both the lattices. However,
this is not the case for $B$ field. Change in the number of components
in the $E$ field corresponds to the change in orientations of the
clusters only whereas change in the number of components in the $B$
field corresponds to the change in the universality class of
DSP. Changing the intensity of the applied fields, a smooth crossover
from DSP to OP, DP and SP is observed. A phase diagram is obtained for
the percolation models under external bias fields. OP, DP or SP could
be obtained as a limiting situation of DSP by changing the field
intensities. The DSP model then can be considered as a most general
model of percolation under external bias fields. The model will be
applicable to the physical situations where crossed directional and
rotational force fields are present in disordered systems.

\vspace{0.5cm}
\noindent
{\bf Acknowledgment:} SS thanks CSIR, India for financial support.

\newpage

\begin{landscape}
\begin{center}

\begin{table}
\begin{tabular}{p{2.4cm}p{3.5cm}p{3.0cm}p{2.30cm}p{2.30cm}p{2.30cm}p{2.30cm}p{2.30cm}}
    \hline
 
    Lattice Type & $E$ field orientation & $d_f$ & $\gamma$ & $\delta$
    & $\eta$ & $\nu_\parallel$ & $\nu_\perp$ \\
     
    \hline 

    Square: & Horizontal\cite{dsp}: & $1.733 \pm0.005$ & $1.85 \pm
    0.01$ & $4.01 \pm 0.04$ & $6.21 \pm 0.08$ & $1.33 \pm 0.01$ &
    $1.12 \pm 0.03$ \\
    & & $1.72\pm0.02$ (FS) &&&&& \\

    & Diagonal: & $1.732 \pm0.006$ & $1.87 \pm
    0.01$ & $4.04 \pm 0.04$ & $6.24 \pm 0.08$ & $1.34 \pm 0.02$ &
    $1.13 \pm 0.02$ \\
    && $1.72\pm0.02$ (FS) &&&&& \\

    Triangular: & Horizontal\cite{sinha}: & $1.775 \pm0.004$ & $1.98
    \pm 0.02$ & $4.30 \pm 0.04$ & $6.66 \pm 0.08$ & $1.36 \pm 0.02$ &
    $1.23 \pm 0.02$ \\
    && $1.80\pm0.03$ (FS) &&&&& \\

   & Semi-diagonal: & $1.777 \pm0.005$ & $2.00 \pm 0.02$ & $4.33 \pm 0.04$
    & $6.71 \pm 0.08$ & $1.37 \pm 0.02$ & $1.24 \pm 0.02$ \\
    & & $1.81\pm0.02$ (FS)  &&&&& \\
    \hline
\end{tabular}
\bigskip
\caption{\label{cls_res} Numerical estimates of the critical exponents
 and the fractal dimension measured for the DSP clusters on the square
 and triangular lattices for different $E$ field orientations. The
 values of the critical exponents and the fractal dimension are found
 independent of the orientation of $E$ field on both the lattices.}
\end{table}
\end{center} 
\end{landscape}

\newpage

\vfill

\vfill

\vfill

\vfill

\vfill

\vfill

\vfill

\vfill
 
\vfill
 
\vfill
 
\vfill
 
\end{document}